\documentclass[a4paper]{jpconf}
\usepackage{graphicx}

\graphicspath{ {nikolaidis-etal-POD_rev1/} }

\usepackage{hyperref}
\hypersetup{
	hyperindex,
	breaklinks,
	colorlinks=true,
	linkcolor=blue,
	citecolor=purple,
	bookmarks=true,
	bookmarksopen=true,
	bookmarksopenlevel=2,
	pdfstartpage={1},
	pdfstartview={FitH},
	pdfview={FitH 0},
	pdfauthor={M-A Nikolaidis, B F Farrell, P J Ioannou},
	pdftitle={The mechanism by which nonlinearity sustains turbulence }
 }

\usepackage{ulem}
\usepackage[sort&compress]{natbib}
\bibpunct{[}{]}{,}{n}{}{,}

\usepackage{microtype}

\usepackage{bm}
\usepackage{doi}
\usepackage{subfigure}

\usepackage{amsmath,amssymb}
\usepackage[usenames,dvipsnames]{xcolor}

\def\U{\bm{\mathsf{U}}}
\def\Uv{\boldsymbol{U} }

\newcounter{saveeqn}%

\newcommand{\be}{\begin{equation}}
\newcommand{\ee}{\end{equation}}
\newcommand{\bdm}{\begin{equation*}}
\newcommand{\edm}{\end{equation*}}
\newcommand{\bea}{\begin{eqnarray}}
\newcommand{\eea}{\end{eqnarray}}

\newcommand{\partialf}[2]
{
 \ifthenelse{\equal{#1}{}}{\frac{\partial}{\partial #2}}{\frac{\partial #1}{\partial #2}}
}

\renewcommand{\(}{\left(}
\renewcommand{\)}{\right)}

\providecommand\bnabla{\boldsymbol{\nabla }}
\providecommand\bcdot{\boldsymbol{\cdot}}

\newsavebox{\astrutbox}
\sbox{\astrutbox}{\rule[-5pt]{0pt}{20pt}}

\usepackage{upgreek}

\renewcommand{\Re}{\textrm{Re}}

\renewcommand{\U}{\boldsymbol{U} }
\renewcommand{\u}{\boldsymbol{u} }

\begin{document}
\title{The mechanism by which nonlinearity sustains turbulence in plane Couette flow}

\author{M-A.~Nikolaidis$^1$, B.~F.~Farrell$^2$, P.~J.~Ioannou$^1$}

\address{$^1$National and Kapodistrian University of Athens, Department of Physics, Panepistimiopolis, Zografos, 15784 Athens, Greece}
\address{$^2$Department of Earth and Planetary Sciences, Harvard University, Cambridge, MA 02138, USA}

\ead{\href{mailto:pjioannou@phys.uoa.gr}{pjioannou@phys.uoa.gr}}

\begin{abstract}
Turbulence in wall-bounded shear flow results from a synergistic 
interaction between linear non-normality and nonlinearity in which non-normal growth of
a subset of  
perturbations configured to transfer energy from the 
externally forced component 
of the turbulent state  to the perturbation 
component maintains the perturbation energy, while
the  subset of energy-transferring perturbations is replenished
by nonlinearity.  Although it is accepted that both linear 
non-normality mediated energy transfer from the forced component of the mean flow and 
nonlinear interactions among perturbations are required to maintain the turbulent state, 
the detailed physical mechanism by which these processes interact in maintaining turbulence
has not been determined.  
In this work a statistical state dynamics based analysis is performed on turbulent 
Couette flow at $R=600$ and a comparison to DNS is used to
demonstrate
that the perturbation component in Couette flow turbulence is replenished
by a 
non-normality mediated
parametric 
growth process in which the fluctuating streamwise mean 
flow has been adjusted to marginal Lyapunov stability.  
It is further shown that the alternative mechanism in which the  
subspace of non-normally growing perturbations is maintained 
directly by perturbation-perturbation nonlinearity does not 
contribute to maintaining the turbulent state.
This work identifies  parametric interaction between the fluctuating streamwise 
mean flow  and the 
streamwise varying perturbations to be the mechanism of the nonlinear 
interaction maintaining the perturbation component of the turbulent state, and identifies the associated 
Lyapunov vectors with positive energetics as the structures of the 
perturbation  subspace supporting the turbulence.
\end{abstract}

\section{Introduction}

Turbulence is widely regarded as the primary exemplar of an essentially nonlinear phenomenon.  
However, the mechanism by which energy is transferred in shear flows from the externally forced component of the flow 
to the broad spectrum 
of spatially and temporally varying perturbations is through  linear  non-normal interaction between these 
flow components~\citep{Henningson-Reddy-1994,Kim-Lim-2000}.  
Nevertheless, nonlinearity participates in an essential way in this cooperative interaction by which turbulence self-sustains.  
Our goal in this study is to provide a more comprehensive understanding of the role of nonlinearity and its interaction with 
linear non-normality in the maintenance of turbulence.

In order to study the mechanism by which nonlinearity between
streamwise varying components participates in
the maintenance of turbulence in shear flow
we begin
by partitioning  the velocity field of  plane parallel Couette flow into streamwise mean and perturbation components,
or equivalently  into the $k_x=0$ and the $k_x \ne 0$ components
of the  Fourier decomposition of the flow field, where $k_x$ is the
wavenumber in the streamwise, $x$, direction. In this decomposition the flow field is partitioned as:
\begin{equation}
\u = \Uv(y,z,t) + \u'(x,y,z,t)~,
\label{eq:mean}
\end{equation}
with cross-stream direction $y$ and spanwise direction  $z$. Note that in this decomposition the mean flow retains
temporal variation in its spanwise structure
as would occur e.g. in the presence of time-dependent streaks.

The non-dimensional Navier-Stokes equations expressed using this mean and perturbation partition are:
\begin{subequations}
\label{eq:NS}
\begin{gather}
\partial_t\U + \underbrace{\U  \bcdot \bnabla  \U }_{N_1}   + \bnabla  P -  \Delta \U /R = \underbrace{- \langle\u ' \bcdot \bnabla  \u '\rangle_x}_{N_2}\ ,
\label{eq:NSm}\\
 \partial_t\u '+  \underbrace{ \U  \bcdot \bnabla  \u ' +
\u ' \bcdot \bnabla  \U }_{N_3}  + \bnabla   p' -  \Delta  \u '/R
= \underbrace{- \( \u ' \bcdot \bnabla  \u ' - \langle\u ' \bcdot \bnabla  \u '\rangle_x \,\)}_{N_4} ~,
 \label{eq:NSp}\\
 \bnabla  \bcdot \U  = 0\ ,\ \ \ \bnabla  \bcdot \u ' = 0\ ,
 \label{eq:NSdiv0}
\end{gather}\label{eq:NSE0}\end{subequations}
where $R= U_w h/ \nu$ is the Reynolds number and $\pm U_w$ the wall velocity at $y=\pm h$.
The  flow satisfies no-slip boundary conditions in the cross-stream direction: $\U (x,\pm h,z,t)= (\pm U_w,0,0)$,
$\u '(x,\pm h,z,t)=(0,0,0)$ and periodic boundary conditions  in the $z$ and $x$ directions. From now on, 
lengths will be normalized with $h$,  velocities with $U_w$, and time with  $h/U_w$. Averaging is denoted with angle brackets $\langle \bcdot \rangle$ with the bracket subscript indicating the averaging variable, so that e.g. the streamwise mean velocity is $\Uv \equiv \langle \u \rangle_x = L_x^{-1} \int_0^{L_x} \u ~dx$, where $L_x$ is the streamwise length of the channel.
The Navier-Stokes equations
with this decomposition will be referred to as the NL system and direct numerical simulations of this system as
NL.  We have indicated 
with an underbrace the terms in the NL system of primary relevance to our study. The advective term in the streamwise mean flow equation~\eqref{eq:NSm} comprising nonlinear interactions among
$k_x=0$ flow components will be referred to as  $N_1$, the Reynolds stress divergence term in~\eqref{eq:NSm}, produced by 
nonlinear interaction between the $k_x$ and $-k_x$ flow components
 with $k_x \ne 0$, will be referred to as $N_2$.  In the perturbation equation~\eqref{eq:NSp} $N_3$ gives the
 linear interaction
 between the  mean flow and  
 the $k_x \ne0$ flow components, which is 
 the source of  the non-normality  giving rise to  the transfer of energy from the mean flow to 
 the perturbations,
 and $N_4$ is  the  nonlinear interactions
 between $k_{x_1}\ne 0$ and  $k_{x_2} \ne 0$, with $k_{x_1}\ne -k_{x_2}$.
 
Transition to and maintenance of a self-sustained turbulent state results
even when only nonlinearity $N_2$ is retained~\cite{Farrell-Ioannou-2012}. By retaining   nonlinearities $N_1$ and $N_2$ 
we obtain the restricted non-linear system (RNL):
\begin{subequations}
\label{eq:RNS}
\begin{gather}
\partial_t\U + \U  \bcdot \bnabla  \U   + \bnabla  P -  \Delta \U /R = - \langle\u ' \bcdot \bnabla  \u '\rangle_x\ ,
\label{eq:RNSm}\\
 \partial_t\u '+   \U  \bcdot \bnabla  \u ' +
\u ' \bcdot \bnabla  \U   + \bnabla   p' -  \Delta  \u '/R
= 0 ~,
 \label{eq:RNSp}\\
 \bnabla  \bcdot \U  = 0\ ,\ \ \ \bnabla  \bcdot \u ' = 0\ .
 \label{eq:RNSdiv0}
\end{gather}\label{eq:RNSE0}
\end{subequations}
It has been confirmed that this RNL system supports a realistic self-sustaining process (SSP) which maintains a turbulent state in
minimal channels~\cite{Farrell-Ioannou-2012,Farrell-Ioannou-2017-bifur}, in channels of moderate sizes at  both low and high Reynolds numbers (at least for $R_\tau \le 1000$)
\cite{Thomas-etal-2014,Farrell-etal-2016-VLSM,Farrell-etal-2016-PTRSA},  and also
in very long channels~\cite{Thomas-etal-2015}.

Consider in isolation the time varying mean flow $\U$ obtained from a state of  turbulence 
either of the NL or of the RNL system.
Sufficiently small perturbations, $\u'$,  on this mean flow evolve  according  to~
\begin{equation}
 \partial_t\u '+  \U  \bcdot \bnabla  \u ' +
\u ' \bcdot \bnabla  \U   + \bnabla   p' -  \Delta  \u '/R~=0~,~~\bnabla  \bcdot \u'=0~,
 \label{eq:RNSp1}
\end{equation}
which is  the perturbation equation~\eqref{eq:RNSp} of the RNL system, 
while in the NL system the finite perturations $ \u'$ obey
the different equation~\eqref{eq:NSp} with the $N_4$ term included. 
In the  self-sustained RNL turbulent state the perturbations, $\u'$, that evolve
under the linear dynamics~\eqref{eq:RNSp} or equivalently under 
\eqref{eq:RNSp1}
remain finite and bounded. Therefore 
the mean-flow, $\U$, of the  RNL turbulent state  is stable in the sense that perturbations, i.e. the streamwise varying flow  components, $\u'$,
that evolve under~\eqref{eq:RNSp1},  have  zero asymptotic growth rate and  the mean flow
can be considered to be in the critical state of neutrality, poised between stability and instability. 
A question that will be addressed in this paper is  whether  the mean flow, $\U$, 
that is obtained from a  simulation of 
NL turbulence  shares this property of being adjusted 
similarly to neutrality in the sense 
that perturbations,  $\u'$,
that evolve under~\eqref{eq:RNSp1}, remain bounded and therefore have  vanishing 
asymptotic growth rate and  the mean flow of the NL system
can therefore be 
considered to be similarly in a
 critical state of neutrality when proper account is taken of dissipation in 
the NL system. If the turbulent mean flow  $\U$ of NL can be shown to be neutral,
in this sense of parametric neutrality, then  the  mechanism of turbulence identified 
analytically in the RNL system, in which the  perturbations arise from parametric instability of the
mean
 flow  with the mean flow being 
  regulated to neutrality through quasi-linear interaction with the
perturbation
field,  will have been extended to NL and the turbulent dynamics of NL
 identified to be essentially the same as that of  the analytically characterized RNL dynamics.
 For this program to succeed it is required to show that the dynamically 
 substantive difference between the RNL system and the NL system, which 
 is the appearance in the NL system of the perturbation-perturbation nonlinearity 
 $N_4$, does not fundamentally change the dynamics of turbulence operating  in the RNL system.

The mechanism determining the statistical state of turbulence in the RNL 
system and its extension to the NL system we are studying 
can be related to an influential conjecture of Malkus: 
``First, that the mean flow will be statistically stable if an
Orr-Sommerfeld type equation is satisfied by fluctuations of the mean; second,
that the smallest scale of motion that can be present in the spectrum of the
momentum transport is the scale of the marginally stable fluctuations of the
mean"~\cite{Malkus-1956}.  In this quote, Malkus  conjectures  that the turbulent mean state is adjusted to a state of neutrality,  
as he defines neutrality, and the turbulent perturbations responsible for the Reynolds stresses are  in the subspace of the neutral 
modes of the Orr-Sommerfeld operator about the mean flow, as he defines it.  We have in common with Malkus 
the concept of adjustment by quasi-linear interaction 
between the mean flow and perturbations as the general mechanism 
determining the statistical state of turbulence in shear flow and we have 
succeeded to show that  RNL turbulence operates with this program.  Extension of the
program to the NL system is the goal of the present work. We differ 
with Malkus in the particulars of this general program, including considering the parametric instability of the time 
dependent streamwise mean state as that being adjusted to neutrality rather than 
the time, spanwise and streamwise mean flow.
While e.g. turbulent convection~\citep{Malkus-1954, Malkus-Veronis-1958} and the
baroclinic turbulence in the midlatitude atmosphere~\citep{Stone-1978}
display a usefully close approximate adherence
to  this conjecture  when both the spatial and temporal means 
are taken to define the mean flow,  the turbulent mean state of wall-bounded
turbulence, defined as the streamwise, spanwise
and temporal mean, $\langle \U \rangle_{z,t}$,
is  hydrodynamically stable and far from neutrality in apparently strong violation of
the Malkus conjecture~\cite{Reynolds-Tiederman-1967}.
However, RNL turbulence suggests that the  program of Malkus~\cite{Malkus-1956} 
was essentially correct and can be extended to wall-turbulence
requiring only the additional recognition that the instability  to be equilibrated is the instability of the
time-dependent operator associated
with linearization about the temporally varying streamwise mean flow, $\U$.

The maximum growth  rate of perturbations to the streamwise 
mean $\U$  that are governed by the linear dynamics of~\eqref{eq:RNSp1} is given by 
the  top Lyapunov exponent
of $\u'$  defined as:
\begin{equation}
\lambda_{Lyap} = \lim_{t \to \infty} \frac{\log |\u'|}{t}~.
\label{eq:exp}
\end{equation}
RNL turbulence with the  definition of mean~\eqref{eq:mean}  satisfies
the Malkus conjecture precisely under our interpretation because for RNL
\begin{equation}
\lambda_{Lyap} = 0 ~.
\label{eq:exp0}
\end{equation}

An issue we wish to examine in this work is  whether NL turbulence (with
the $N_4$ term included) is
similarly neutral in the Lyapunov sense with its perturbations being similarly supported by parametric 
growth on its fluctuating mean flow with perturbation structure being  
associated with the predicted Lyapunov vector structure.
Specifically,  we mean whether fluctuations, $\u'$, evolving 
under~\eqref{eq:RNSp1}  on the  time dependent mean flow, $\U$, 
that has been obtained from a turbulent simulation
of the NL equations, have $\lambda_{Lyap}$, as defined in~\eqref{eq:exp}, zero 
when proper account is taken of dissipative processes  and  the 
predicted Lyapunov structure can be verified to be maintaining the perturbations.
We caution the reader 
that the Lyapunov exponents we are calculating  are not  the
familiar Lyapunov exponents of small perturbations from a turbulent
trajectory, which  is associated with the growth  of
perturbations $\delta \U$, $\delta \u'$  to the tangent linear
dynamics of the full NL system linearized about a turbulent
trajectory $\U$, $\u'$. The turbulent trajectory  $\delta \U$, $\delta \u'$
is chaotic and will have  typically many positive Lyapunov 
exponents. We instead  
calculate only the Lyapunov exponents and structures of perturbations, $\u'$, evolving
under the linear dynamics~\eqref{eq:RNSp1} with time dependent  mean flow $\U$.
It is important to recognize that 
the parametric perturbation evolution equation~\eqref{eq:RNSp1} governing
 the perturbation Lyapunov vector dynamics of RNL 
 is not limited to small perturbation amplitude
 and the nonlinearity required to regulate the perturbations to their finite statistical 
 equilibrium state, although not present in~\eqref{eq:RNSp1}, is contained
  in  the Reynolds 
 stress feedback term $N_2$ appearing in the mean equation. 

Choice of the mean  used in the cumulant expansion is 
fundamental to formulating an SSD to gain an understanding 
of the mechanism of turbulence in shear flow.   SSD analysis
reveals that  transition to and maintenance of turbulence
occurs in association with the breaking of statistical symmetries
of the laminar state through a sequence of bifurcations.
The laminar state in Couette flow
has spanwise, streamwise and temporal statistical homogeneity.
When the SSD of Couette flow is closed at second order it has been shown that   at a first
critical Reynolds number the spanwise symmetry is broken,  and at a second
higher Reynolds number the temporal statistical homogeneity
is  broken, coincident with transition  to the turbulent state
\citep{Farrell-Ioannou-2012,Farrell-Ioannou-2017-bifur}.
 The centrality of spanwise variation of the mean flow, which is associated with 
the streak component, to the maintenance of turbulence
has been  demonstrated by numerical experiments that show
turbulence is not sustained when the streaks are sufficiently
damped or removed~\cite{Jimenez-Pinelli-1999}.
Given that in  the SSD turbulent state the spanwise and temporal homogeneity
are broken,   it is dynamically inconsistent to use  
either the spanwise or the temporal mean to
separate the flow field into mean and perturbation components. It is necessary 
to allow both spanwise and temporal variations in the mean structure
which requires representation~\eqref{eq:mean}.

It  remains an open question  whether in the turbulent state the streamwise statistical homogeneity is also broken.
The evidence that is usually presented to support  breaking of the  streamwise statistical homogeneity
is that no arbitrarily long streamwise  structures  are observed in simulations, but
this evidence is based on individual realizations which may not  reflect the symmetries of the statistical state.
The evidence supporting the non-breaking of the streamwise symmetry is that RNL 
has been shown to support a realistic turbulent state under the assumption
of a streamwise mean ($k_x=0$) in long turbulent channels,
 demonstrating  that there exists  a statistical turbulent state with statistical streamwise homogeneity~\citep{Thomas-etal-2015}.

\begin{figure*}
\centering\includegraphics[width=24pc]{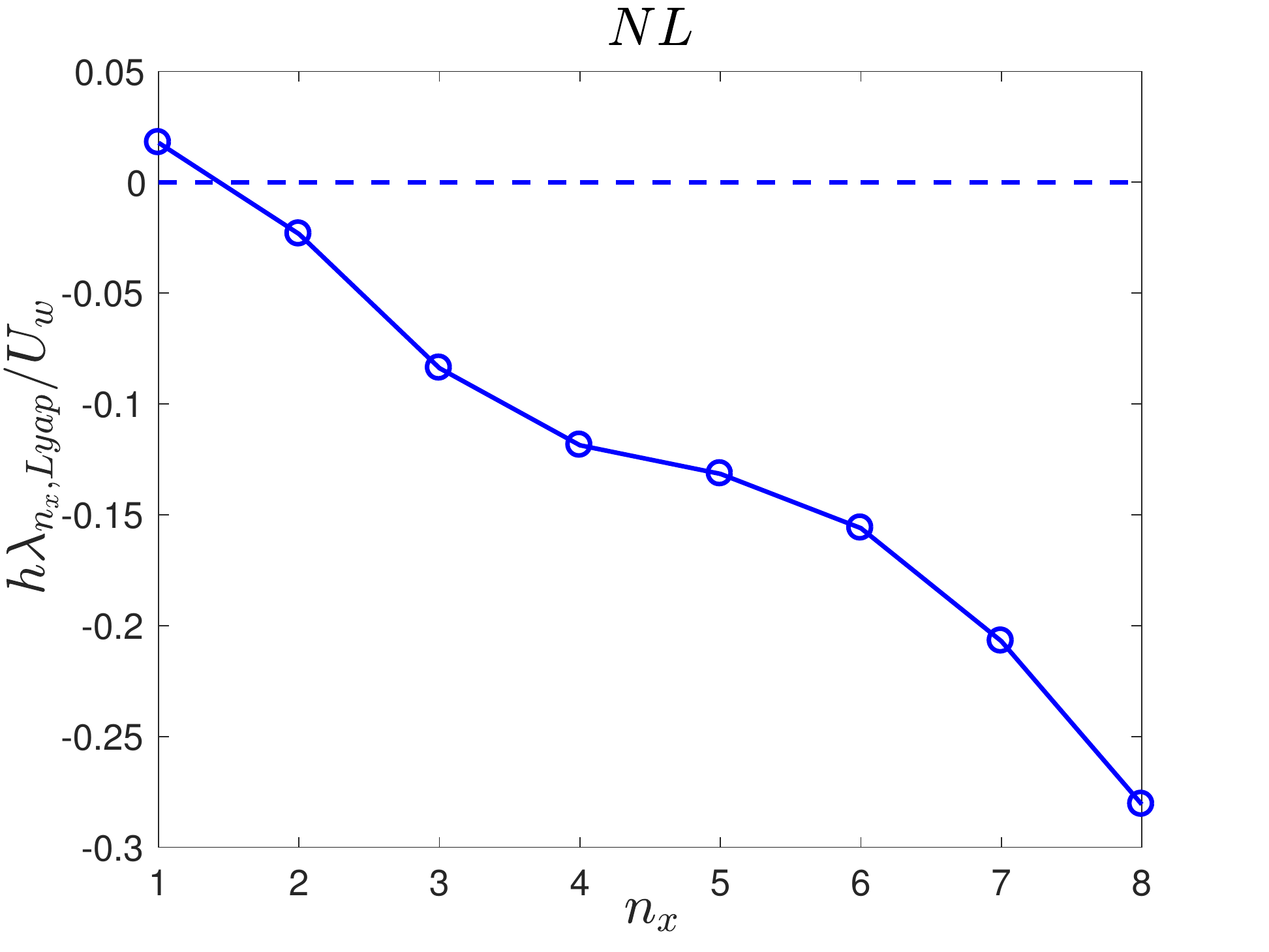}
\vspace{-1em}
\caption{
The top Lyapunov exponent of perturbations  with
channel wavenumbers $n_x=1,\dots, 8$ evolving under the time-dependent 
turbulent mean flow in NL, $\U$.
The Lyapunov exponent of all $n_x  \ge 2$ components of $\u'$
is negative.  For comparison, the 
least stable mode of the  streamwise-spanwise-temporal mean of $\U$ has 
decay rate $\sigma=-0.12 ~U_w/h$   at $n_x=1$ and  $n_z = 3$. The Lyapunov 
exponent was nondimensionalized  using advective time units, $h/U_w$. A 
plane Couette channel at $R = 600$ was used.} \label{fig:Lyap_k}
\end{figure*}

Comparing RNL and NL dynamics provides a new perspective on the
role of the perturbation-perturbation nonlinearity $N_4$ in the maintenance and regulation of 
turbulence.
The $N_4$ term  in~\eqref{eq:NSp}
does not contribute directly to maintaining the perturbation energy because  the  perturbation-perturbation interactions
redistribute energy  internally among the streamwise $k_x \ne 0$ components of the flow and
the term $ \langle\u' \bcdot N_4\rangle_{x,y,z}$ is zero\footnote{In our simulations time discretization produces a $ \langle\u' \bcdot N_4 \rangle_{x,y,z,t}$ of the order of  $-0.0005 U_w^3/h$ which provides an error estimate for the 
accuracy of our results.}.
Consequently, from~\eqref{eq:NSp} we obtain that the perturbation energy density, $E_p = \langle |\u'|^2 / 2\rangle_{x,y,z}$, evolves according to:
\begin{equation}
 \frac{d E_p}{dt} =    \underbrace{\left \langle \u' \bcdot \left ( -  \U  \bcdot \bnabla  \u ' -
\u ' \bcdot \bnabla  \U   +   \Delta  \u '/R \right ) \right  \rangle_{x,y,z}}_{\dot E_{linear}}~.
 \label{eq:Ep1}
\end{equation}
just as in RNL turbulence. The  term  $\dot E_{linear}$ gives   the energy transfer
to the streamwise-varying perturbations by  interaction with the mean $\U$. The Lyapunov exponent of the 
$\u'$ associated with the  mean flow, $\U$, taken from
an NL simulation, and defined in~\eqref{eq:exp}  is also given by
the time-average of the instantaneous  energy growth rates:
\begin{equation}
\lambda_{Lyap} = \left \langle    \frac{1}{2 E_p} \frac{d E_p}{dt} \right \rangle_{t}~.
\label{eq:mle}
\end{equation}
Equation~\eqref{eq:mle} gives  the top Lyapunov exponent of the NL  perturbation dynamics, but  a full spectrum of exponents can also  be obtained using
orthogonalization techniques~\cite{Farrell-Ioannou-1996b}.  

This top Lyapunov exponent should be contrasted with the  exponent obtained
by inserting into~\eqref{eq:Ep1} and~\eqref{eq:mle}  the $\u'$ taken from a simulation of
the NL turbulent state.   This $\u'$ is bounded 
because it is the perturbation state vector
and therefore this exponent   is exactly $\lambda_{state}=0$. While only the top Lyapunov 
vector is excited in RNL, in NL a spectrum of Lyapunov 
vectors comprise the $\u'$  of the NL state and because 
these are orthogonal in Fourier representation it is useful to consider the energetics 
of each of the streamwise Fourier components of $\u'$ separately.
 If we decompose the perturbation field into its streamwise components:
\begin{equation}
\u'  = \sum_{n_x=1}^N ~\u'_{n_x} (y,z,t)  e^{ i k_x x}~,
\end{equation}
with $k_x= 2 \pi n_x /L_x$,  the effective growth rate of 
each of the streamwise components $\u'_{n_x}$ will also be individually zero.  This requires 
that
\begin{equation}
\lambda_{n_x,  state} = \left \langle   \frac{{\rm Re }\langle \u'_{n_x} \bcdot \left ( -  \U  \bcdot \bnabla  \u '_{n_x} -
\u '_{n_x} \bcdot \bnabla  \U   +   \Delta  \u '_{n_x} /R +N_{4, n_x} \right ) \rangle_{y,z}}{ 2 E_{n_x}}  \right \rangle_{t} = 0.
\label{eq:Ep2s}
\end{equation}
 where $E_{n_x} = \langle |\u'_{n_x}|^2 / 2\rangle_{y,z}$ is the kinetic energy of the $n_x$ streamwise component and $N_{4, n_x}$ is the $n_x$  
 streamwise component of  the $N_4$ term  in~\eqref{eq:NSp} (${\rm Re}$ denotes the real part of a quantity).   In the energetics of NL  in addition to 
 the rate of energy transfer  to the perturbations from the mean flow:
\begin{equation}
\dot E_{def, n_x} = {\rm Re } \left  \langle \u'_{n_x} \bcdot \left ( -  \U  \bcdot \bnabla  \u '_{n_x} -
\u '_{n_x} \bcdot \bnabla  \U    \right ) \right  \rangle_{y,z}~,
\label{eq:edef}
\end{equation}
and perturbation energy dissipation rate:
\begin{equation}
\dot E_{dissip, n_x} = \frac{1}{R}  {\rm Re} \left   \langle \u'_{n_x} \bcdot  \Delta  \u '_{n_x} \right  \rangle_{y,z}~, 
\end{equation} 
which are the only terms  in~\eqref{eq:Ep1} that are involved in the determination of the Lyapunov exponent, there is the additional term
\begin{equation}
\dot E_{nonlin,n_x} =  {\rm Re} \left   \langle \u'_{n_x} \bcdot  N_{4, n_x} \right  \rangle_{y,z}~, 
\end{equation} 
that is only present in the NL equations and specifies the net energy transfer rate to the other nonzero streamwise
components.

\begin{figure*}
\centering\includegraphics[width=30pc]{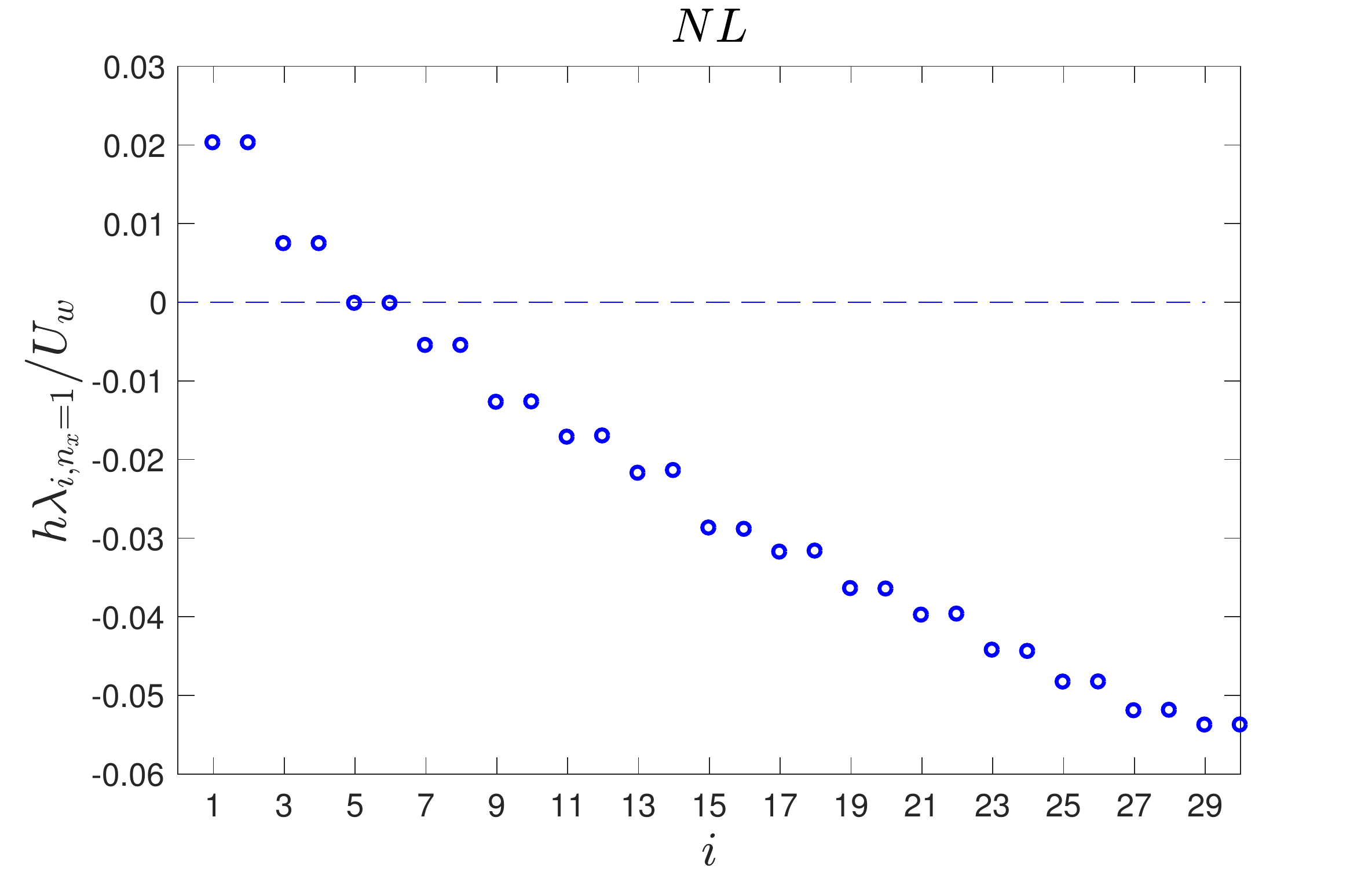}
\vspace{-1em}
\caption{The first 30 Lyapunov exponents of  $n_x=1$ perturbations
to the NL turbulent flow.
Note that each exponent corresponds to two states that allow for streamwise translation of the Lyapunov vector.} \label{fig:lyaps1}
\end{figure*}

The dynamical significance  of  $N_4$ in sustaining the turbulent state is revealed  by comparing
the perturbation energetics  under the influence of  the NL mean flow $\U$ with and without the term $N_4$.
This can be achieved  by calculating  the Lyapunov exponents $\lambda_{Lyap}$ of  
the $\U$ obtained from an NL simulation and the associated Lyapunov vectors 
together with  the   contributions of each
to the growth rate from $\dot{E}_{def,n_x}$,   $\dot{E}_{dissip,n_x}$, $\dot{E}_{nonlin,n_x}$  and  comparing these rates to those obtained
when the term $N_4$ is included.  Although the $N_4$ term   
is energetically neutral   it
may have a profound impact on the energetics by modifying  the perturbations to
extract more or less energy from the
mean flow. 
 If the term $N_4$ is not fundamental to sustaining the turbulence and regulating the mean flow  
 but instead the mechanism of RNL is fundamentally responsible for maintaining NL turbulence 
 the following three conditions should be satisfied: 
 {(\emph i)}
 the top Lyapunov exponent, $\lambda_{Lyap,n_x}$,  associated with streamwise components
$n_x$ of the turbulent field should be  neutral after accounting for the transfer of energy
to the other streamwise perturbation components, which would indicate
that the turbulent state has been regulated to 
neutralize the top Lyapunov vector growth rate (maximum Lyapunov  exponent)
 coincident with the (necessary) neutralization of the state vector, 
  ($\emph ii$) the transfer of energy from the mean flow by the top Lyapunov vector
   $\dot{E}_{def,n_x}$  should exceed that by the state 
  vector indicating that $N_4$ has disrupted the Lyapunov vector 
  making it less effective at transferring energy from the mean 
  flow, and ($\emph iii$) the Lyapunov vectors
span the energy and the energetics of the NL perturbation field in a 
convincingly efficient manner, 
most tellingly if they span it in the order of their growth rate.  
Satisfying these conditions 
would  
strongly support the conclusion that the turbulence 
is being maintained  primarily through the parametric perturbation growth 
process associated with the temporal variation of $\U$ that supports RNL, 
without substantial contribution from the $N_4$ nonlinearity.
The 
alternative is that the $N_4$ term has a first-order effect on the energetics which would imply centrality in the dynamics of turbulence of
the alternative role for $N_4$ which is to   replenish the subset of perturbations lying in the directions of growth.
This distinction in mechanism can be clarified by observing that, if the mean flow is chosen to be
the time-independent streamwise-spanwise-temporal mean,  which in a boundary layer flow is the stable Reynolds-Tiederman profile~\cite{Reynolds-Tiederman-1967},
the $N_4$ nonlinearity must assume this role if turbulence is to be 
sustained as the parametric mechanism is not available. Turbulence could be in principle
sustained by this mechanism if $N_4$ were  
sufficiently effective in scattering  perturbations back into the 
directions of non-normal growth as is commonly hypothesized  in toy models
\citep{Trefethen-etal-1993,Gebhardt-Grossmann-1994,Baggett-Trefethen-1997,Grossmann-2000}.  
However, it is known that this is not the case~\citep{Jimenez-Pinelli-1999}.

\begin{figure*}
\centering\includegraphics[width=30pc]{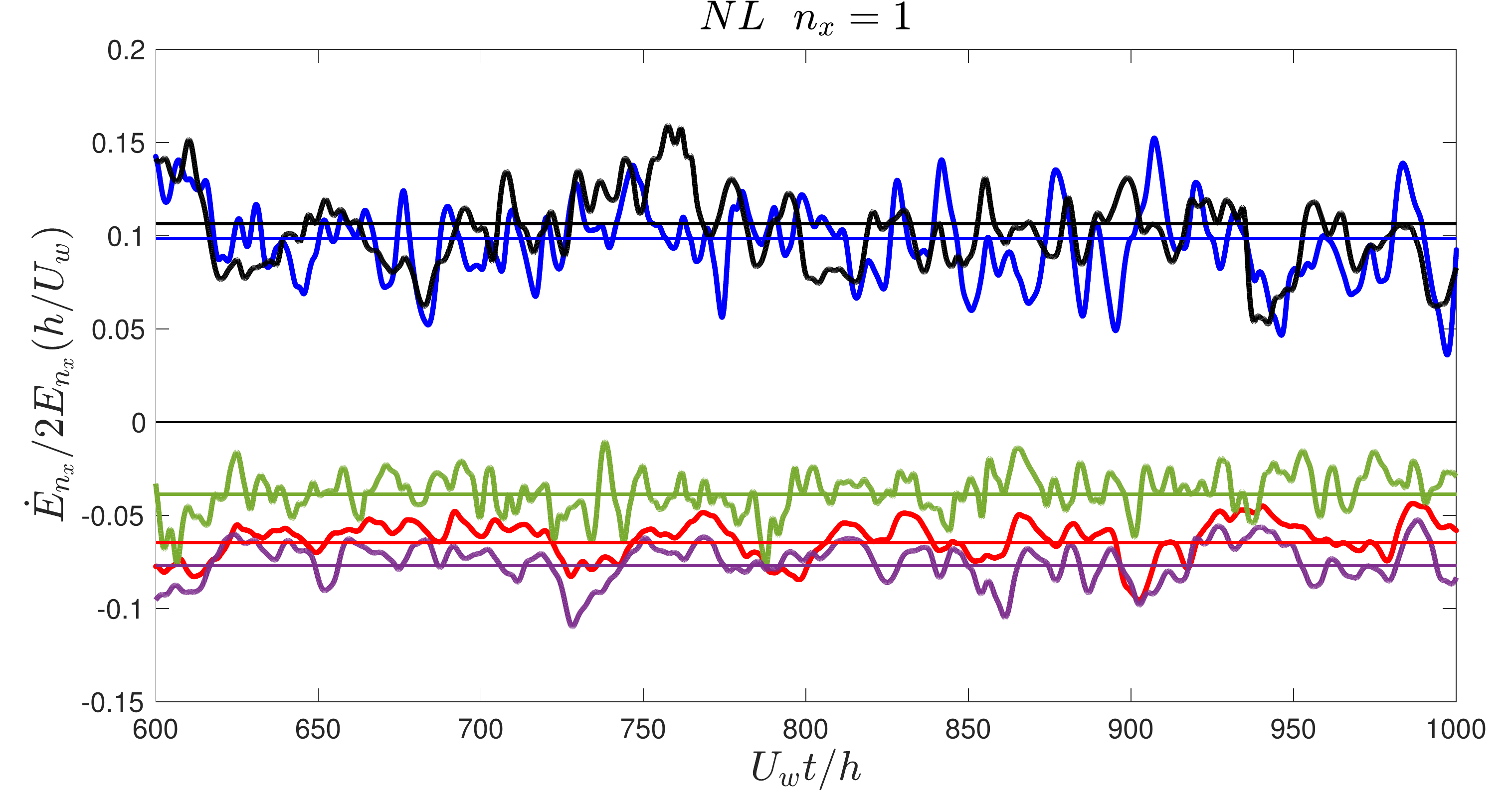}
\vspace{-1em}
\caption{Contribution to the instantaneous energy  growth rate of the $n_x=1$ perturbation component of the NL simulation: extraction 
from the fluctuating $n_x=0$ mean component $\dot E_{def, n_x}/(2 E_{n_x})$ (blue, solid); 
loss to dissipation $\dot E_{dissip, n_x}/(2 E_{n_x})$ (red, solid); transfer to the other $n_x >1$ streamwise 
components $\dot E_{nonlin,n_x}/(2 E_{n_x})$ (green).
The mean values of these rates are indicated with the dashed lines with the corresponding color. These average rates sum to $\lambda_{state}=0$.
The corresponding rates for the first
Lyapunov vector
are shown in black and purple dash-dotted lines (there is no energy transfer to the other components as $N_4$ is absent in this calculation). 
These rates sum to the
Lyapunov exponent $\lambda_{Lyap} =0.02 U_w/h$, which is comparable  to $\dot E_{nonlin,n_x}/(2 E_{n_x})$. }
 \label{fig:comp1}
\end{figure*}


\section{The Lyapunov exponent of the mean flow in Couette turbulence  at $R=600$}


\begin{center}
\begin{table}
\caption{\label{table:geometry}The channel is periodic in the streamwise, $x$, and spanwise, $z$, direction
and at the channel walls $y=\pm h$ the velocity is $\u = (\pm U_w, 0, 0)$. The channel length is  $L_x$ and $L_z$  in the streamwise and spanwise directions respectively.
The number of streamwise and spanwise Fourier components is  $N_x$ and $N_z$
after dealiasing  in the streamwise and spanwise direction by the $2/3$ rule,
and we use  $N_y$ grid points in the wall-normal direction.
$R = U_w h / \nu$ is the Reynolds number of the simulation, with $\nu$ the kinematic viscosity.
}
\centering\vspace{.8em}
\begin{tabular}{@{}*{5}{c}}
\break
 Parameter  & $[L_x,L_z]/h$ &$N_x\times N_z\times N_y$& $R$ \\
 NS600   & $[1.75\pi\;,\;1.2\pi]$&$17\times 17\times 35$&600   \\
\end{tabular}
\end{table}
\end{center}

\vskip-0.2in
Consider a Couette turbulence simulation at $R=600$ in a  periodic channel with parameters given in Table \ref{table:geometry}.
This is  a larger channel compared to the minimal Couette flow that was studied by 
Hamilton, Kim \& Waleffe~\citep{Hamilton-etal-1995} at $R=400$.
RNL turbulence with these parameters at $R=600$
was systematically examined recently
\citep{Farrell-Ioannou-2017-sync}.
%

We first calculate the
Lyapunov exponent $\lambda_{Lyap}$
of the NL mean flow by estimating~\eqref{eq:mle} from a long integration  of~\eqref{eq:RNSp1} with the mean flow $\U$  obtained from
a turbulent NL simulation.
The initial state $\u'$ is inconsequential
because,  with  measure zero exception,  any random initial condition converges 
in this system with exponential accuracy
to the structure associated with the largest  Lyapunov exponent. The full spectrum of Lyapunov exponents can be obtained by an
orthogonalization
procedure.
For a  discussion of the calculation and properties of
Lyapunov exponents and the structures associated with them  refer to Refs.~\citep{Farrell-Ioannou-1996b,Farrell-Ioannou-1999, Farrell-Ioannou-2017-sync, Wolfe-Samelson-2007,Cvitanovic-etal-2016}. Because of the streamwise
independence of $\U$,  the different streamwise Fourier components of $\u'$ in this Lyapunov exponent calculation, in which the
 $N_4$ term is absent, evolve
independently and the structure (the Lyapunov vector) associated with a given Lyapunov exponent
has streamwise structure confined to a single  streamwise  wavenumber $k_x = 2 \pi n_x  h/ L_x$, 
corresponding to  the $n_x$  streamwise wavenumber.

The top Lyapunov exponent  at each $n_x$ is shown in figure~\ref{fig:Lyap_k}. This
plot  reveals 
 that the time dependent streamwise mean flow 
$\U$  is  asymptotically stable to all perturbations with $n_x>1$ with only  the $n_x=1$ streamwise component  supporting
a positive Lyapunov exponent of
$\lambda_{Lyap}\approx 0.02 U_w/h$.  
  Recall that in RNL the top Lyapunov exponent also has wavenumber  
  $n_x=1$ and is exactly 
  zero  consistent with  mean $\U$ being adjusted by the feedback 
  through the Reynolds stress term $N_2$ 
to   exact  neutrality.  The top Lyapunov exponent  obtained using  the $\U$ of  NL 
must be  positive only to the degree required to account for 
the  energy  exported to other perturbations.
The degree of  positivity  of $\lambda_{Lyap}$ is necessarily adjusted by feedback 
through $N_2$ to account for  this loss. With this consideration  the NL mean flow $\U$
can be verified to be neutral. 

Strictly speaking, the $n_x=1$ component is associated with a  pair of Lyapunov exponents
corresponding to the real and imaginary part of a single structure.  This
degeneracy of the Lyapunov exponents results from the streamwise homogeneity of the linear
operator which implies that the streamwise phase of the Lyapunov vectors is arbitrary.
The growth rates of  the  first  30 Lyapunov exponents associated with  $n_x=1$ are  shown in figure~\ref{fig:lyaps1}.
Specifically, the perturbation subspace
is spanned by a set of Lyapunov vectors comprised of  pairs of structures sharing a single spanwise/cross-stream structure 
but having respectively $\sin (k_x x)$ and $\cos(k_x x)$ streamwise dependence.
From figure~\ref{fig:lyaps1} it is apparent that
at this Reynolds number and for this minimal  channel the NL turbulent mean flow supports
 two positive  Lyapunov exponent pairs.

\begin{figure}
 \centering
\begin{subfigure}
 \centering\includegraphics[width=2in]{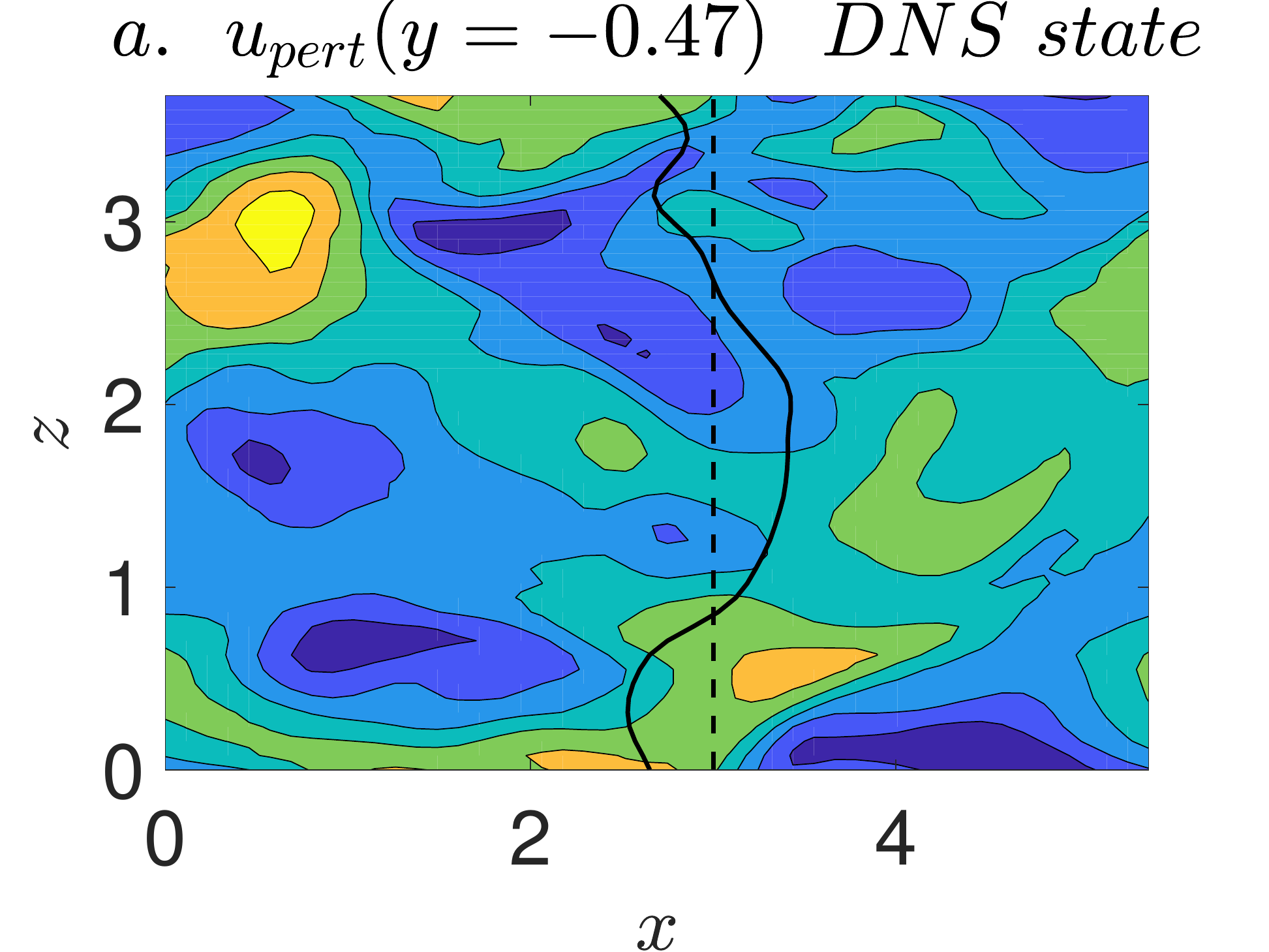}
\end{subfigure} %
\begin{subfigure}
 \centering\includegraphics[width=2in]{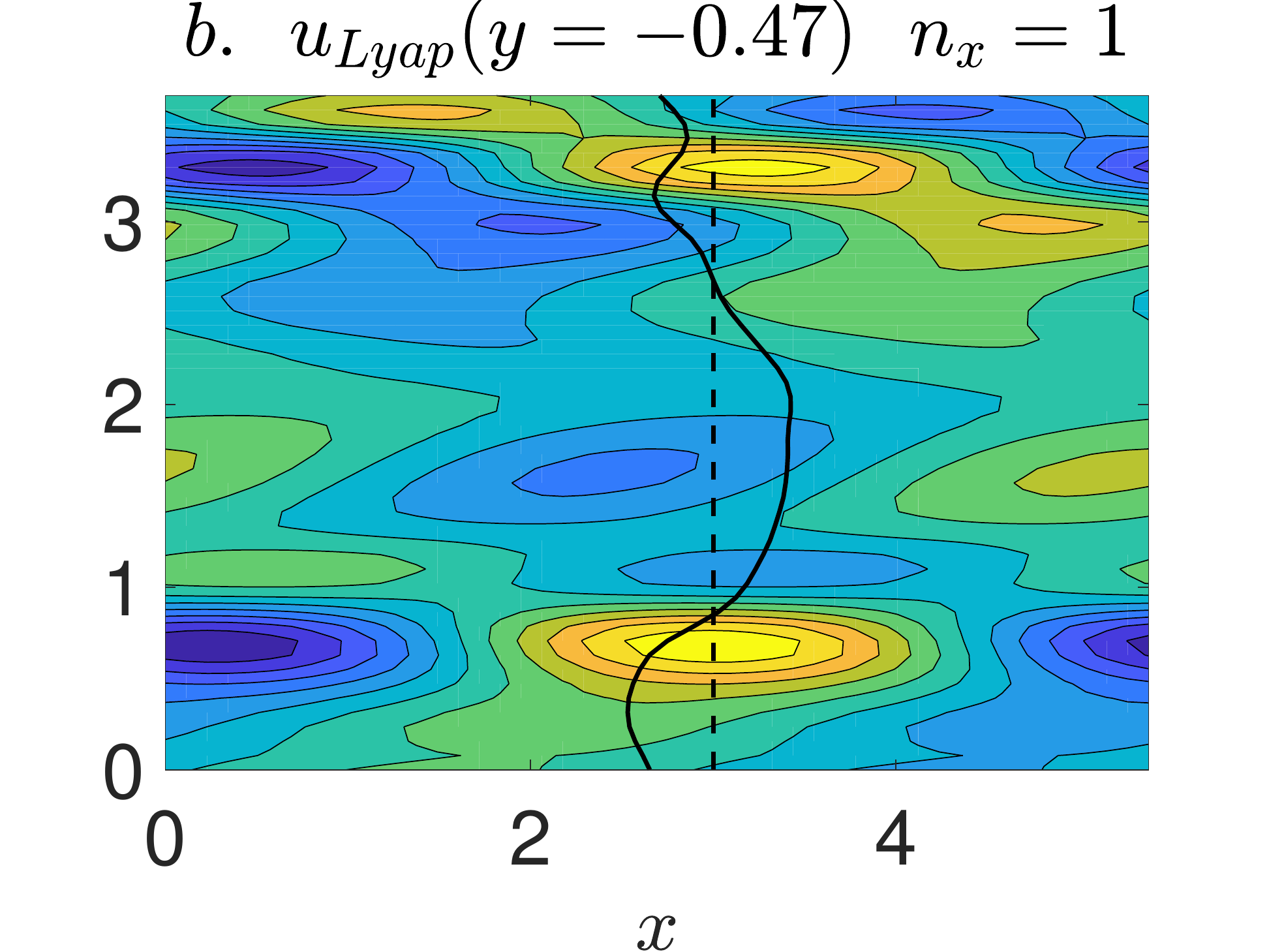}
\end{subfigure}%
\begin{subfigure}
 \centering\includegraphics[width=2in]{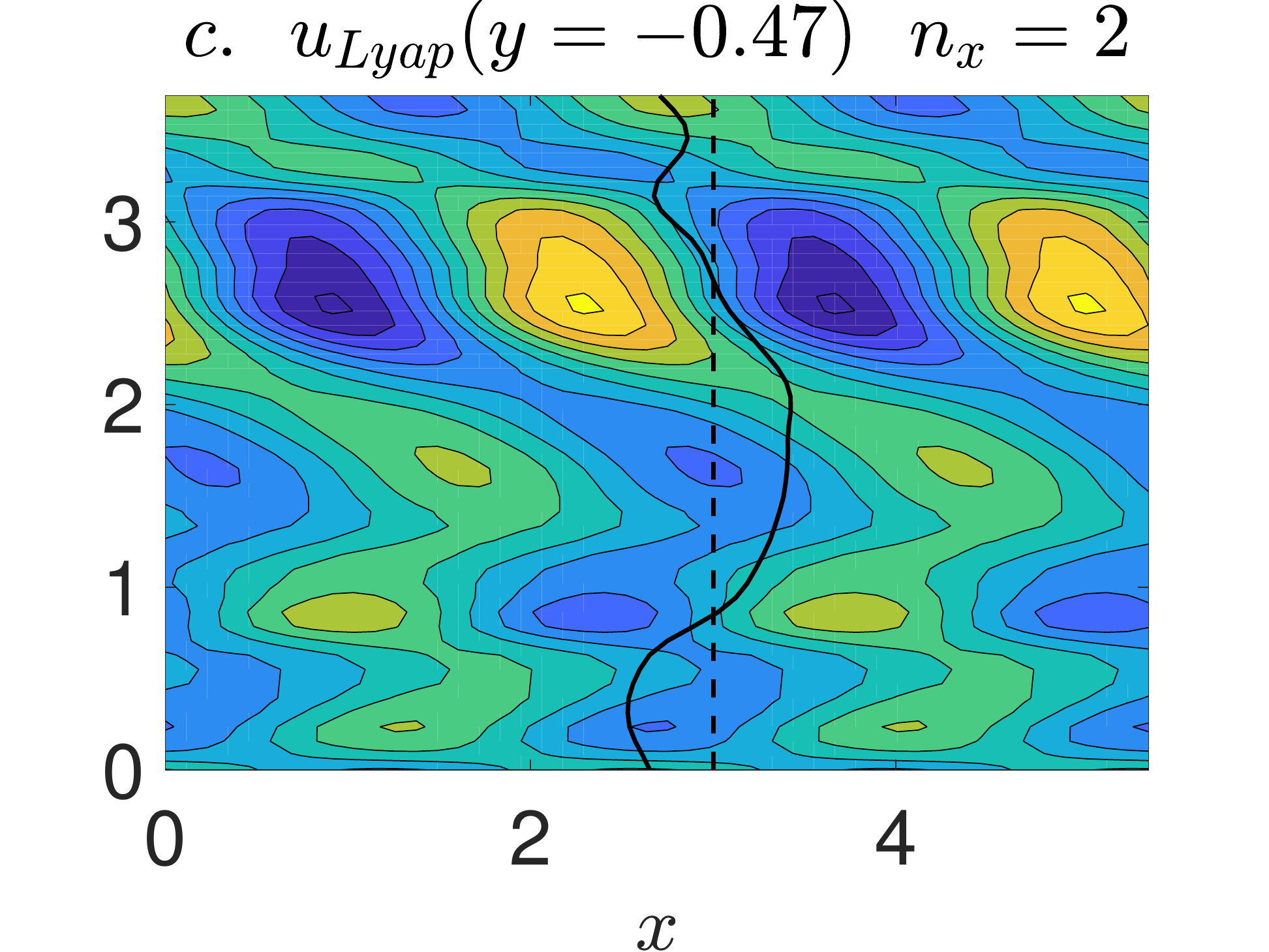}
\end{subfigure}
\caption{\label{fig:snap} Snapshots at the same time showing
the  streamwise perturbation velocity $u'$  of the
NL state (panel (a)), of  the top Lyapunov vector which has streamwise channel wavenumber $n_x=1$
 (panel (b)) and
the decaying top Lyapunov vector associated with $n_x=2$ (panel (c)). Also shown is  the streak component of the streamwise mean
flow $U - \langle U \rangle_z $ (black line).}
\end{figure}


\begin{figure*}
\centering\includegraphics[width=30pc]{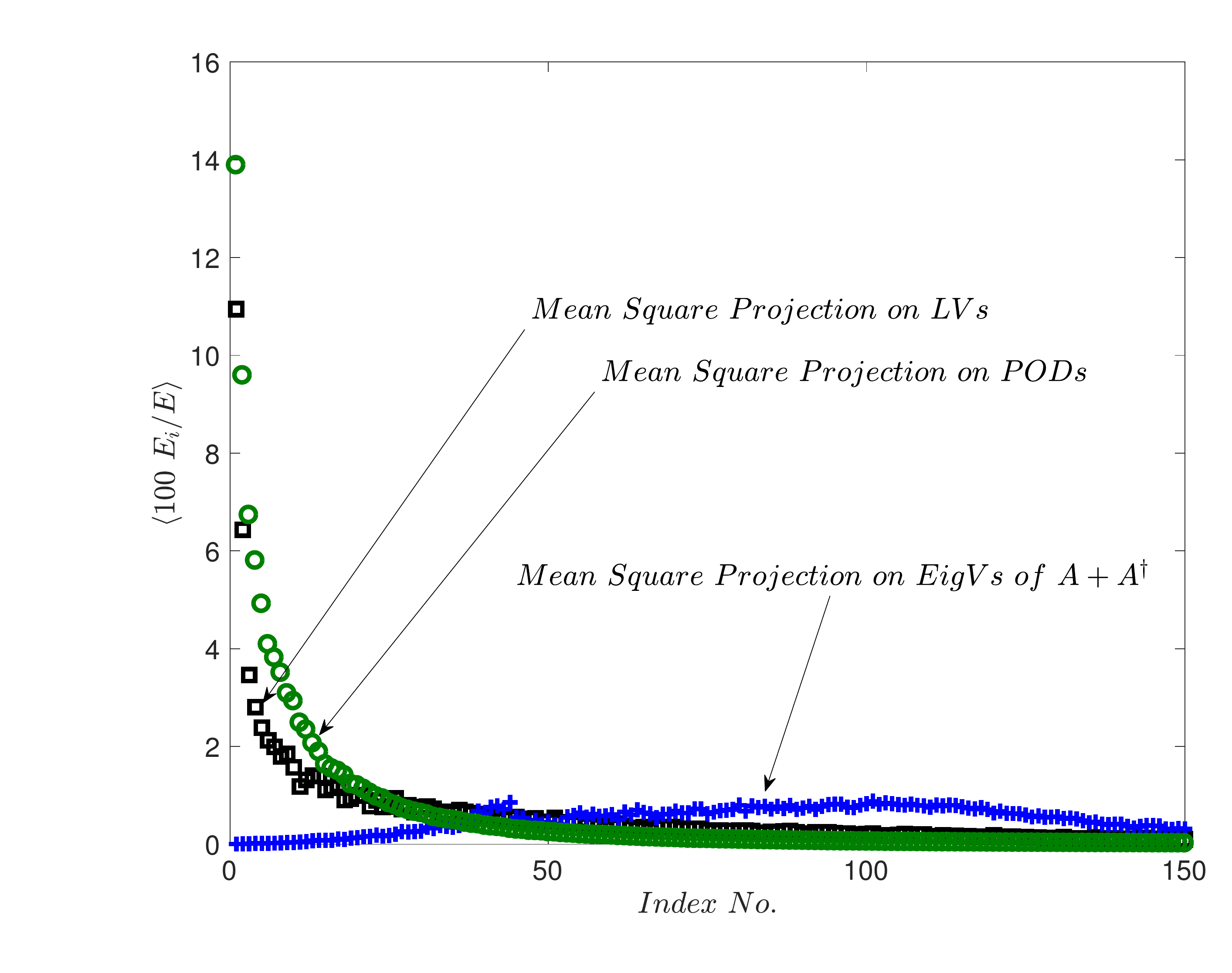}
\vspace{-1em}
\caption{Average energy percentage of the $n_x=1$ flow accounted for by each pair of Lyapunov vectors (black squares). 
Average energy percentage of the $n_x=1$ flow accounted for by the eigenvectors of  $ A+A^{\dagger}$ 
ordered in descending order of their eigenvalue (blue crosses). 
$A$ is the  operator in 
\eqref{eq:RNSp1} governing the linear evolution of the perturbations about $\U$.
The eigenvectors of $ A+A^{\dagger}$ 
are the orthogonal directions of stationary  instantaneous perturbation energy growth rate, with this 
 growth rate 
given by the corresponding eigenvalue. The perturbation component of the turbulent flow 
is adjusted so as to have a small projection on the first eigenvectors of $A+A^{\dagger}$
 associated with large instantaneous energy growth rates.  
 Also shown is the energy percentage accounted for by the PODs of the $n_x=1$ component
 of the NL perturbation state (green circles).
 } \label{fig:proj}
\end{figure*}

\begin{figure*}
\centering\includegraphics[width=28pc]{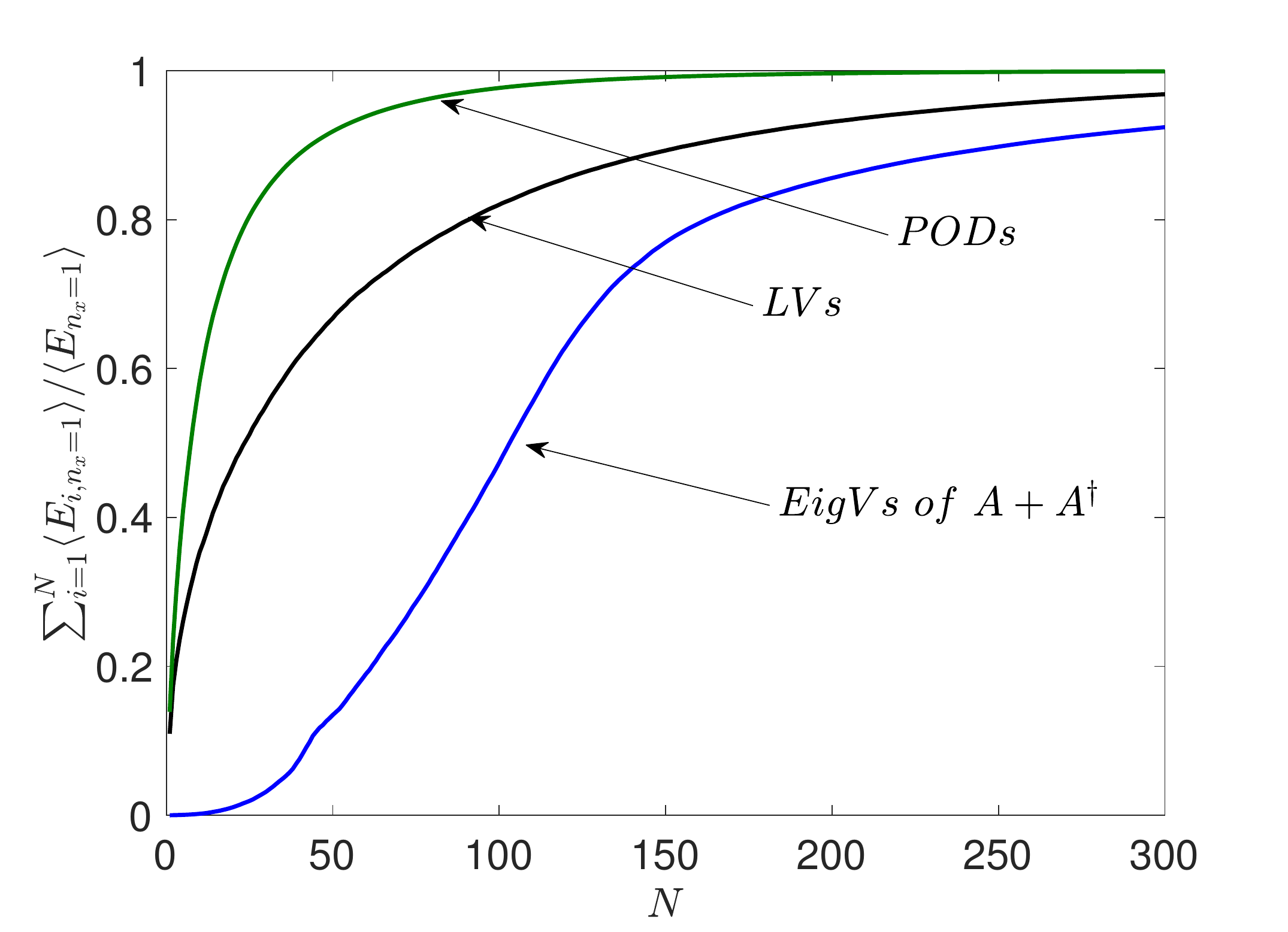}
\vspace{-1em}
\caption{ Average fraction of the energy  of the $n_x=1$ flow accounted for by the first $N$ pairs of Lyapunov vectors (black),
by the first  $N$ pairs of  eigenvectors of  
of  $A+A^{\dagger}$ (blue)  ordered in descending order of their eigenvalue, 
and by the first $N$ pairs of POD's of the $n_x=1$ component of the NL state (green). 
} \label{fig:cum}
\end{figure*}

Contributions to the Lyapunov exponent 
from  mean flow energy transfer and from dissipation
are plotted as a function of time in figure~\ref{fig:comp1}.  The growth rate associated with energy transfer
from the fluctuating streamwise  mean is on average $0.1066 U_w/h$, while the dissipation rate is on average
$0.0862 U_w/h$ resulting in the positive  Lyapunov exponent $\lambda_{Lyap} = 0.0204 U_w/h$.
These transfers occur
when perturbations evolve under the dynamics of the fluctuating streamwise mean flow $\U$ associated with NL turbulence but 
in the absence
of two effects: ($\emph i$) 
disturbance to the perturbation structure by the perturbation-perturbation nonlinearity $N_4$ 
and ($\emph ii$) transfer of energy to other perturbations by $N_4$.  
This result demonstrates that the parametric growth mechanism is able to 
maintain the perturbation turbulence component against dissipation 
with sufficient additional energy extraction to account for transfer to the other scales.  
The last can 
be roughly estimated by the energy transfer rate from $n_x=1$ to $n_x\ge 2$ in the full NL 
simulation, which is on average $-0.0386 ~U_w/h$ or $0.8 u_{\tau}/h$ in terms of  friction velocity, as  shown in Fig. 3.
%




\begin{figure*}
\centering\includegraphics[width=30pc]{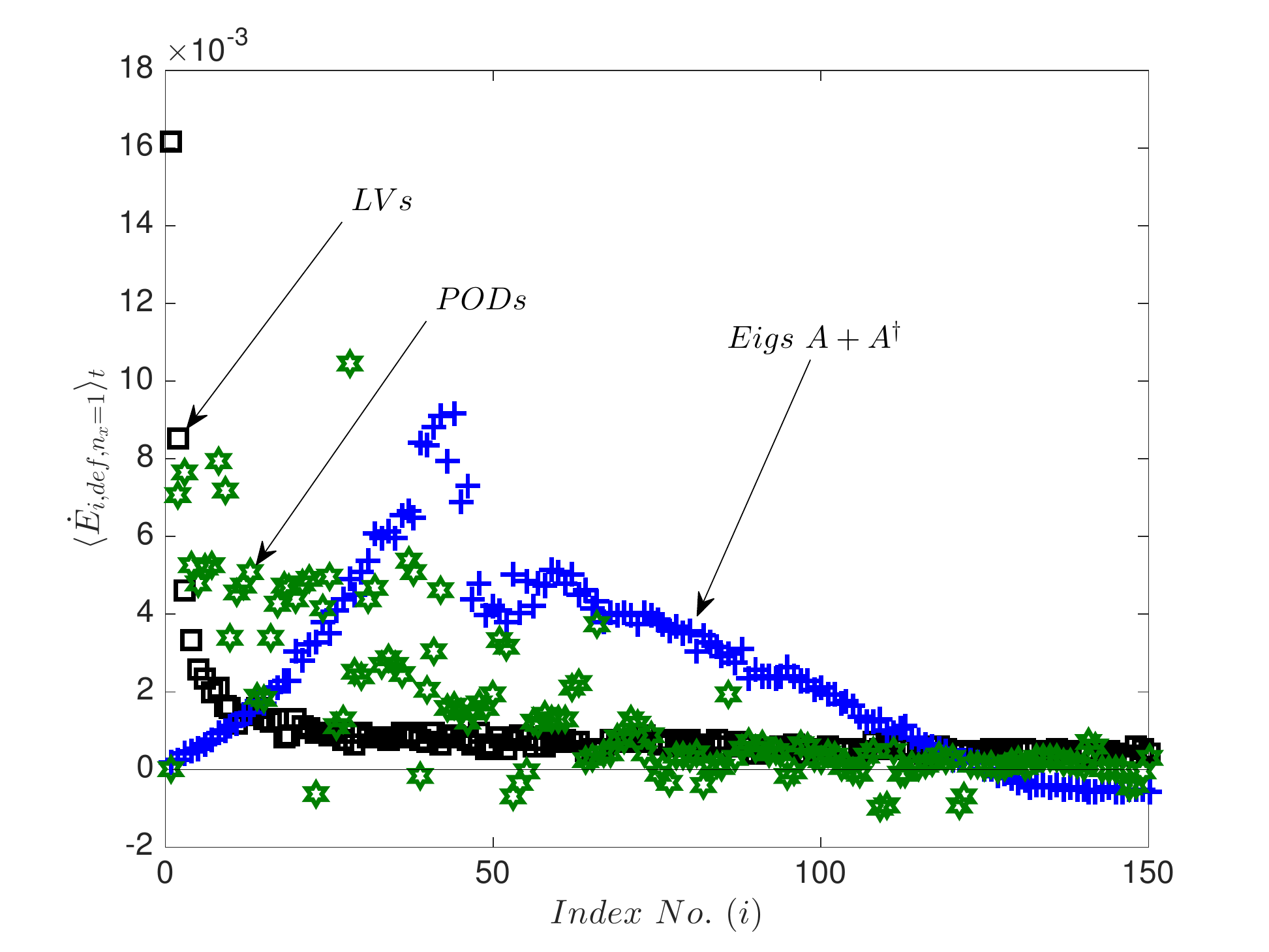}
\vspace{-1em}
\caption{  Contribution of the  Lyapunov vectors (black squares), the PODs (green stars) and  the eigenvectors
of $A+A^\dagger$ (blue crosses) to the energy extraction rate, $\dot E_{def}$,   from the mean flow. 
 } \label{fig:edot}
\end{figure*}


We now contrast the energetics of the Lyapunov vectors on the turbulent NL mean flow just shown
with the corresponding energetics of the
 $n_x=1$ Fourier component of the
state vector obtained from the turbulent NL simulation itself  in order to  determine
whether the $N_4$ term has the effect of influencing the perturbations to be in a more or less
favorable configuration for extracting energy from the mean flow.
These results are also shown in figure~\ref{fig:comp1} from which it can be seen that
the transfer from the mean $\U$ with the influence of the $N_4$ term included
is slightly less  than that  achieved by the first Lyapunov vector in the absence of the influence of $N_4$:  the transfer to the
NL state produces  growth rate $0.0985 U_w/h$  compared to  $0.1066 U_w/h$
for the case of the unperturbed  Lyapunov vector on the NL mean flow.
This demonstrates that the nonlinear term does not configure the flow states so as to extract more energy from
the mean.   However, despite the fact that the energy extracted from the mean flow by the NL perturbation state and 
the  first Lyapunov vector are nearly equal when averaged over time, the correlation coefficient of their time series,  
shown in figure~\ref{fig:comp1}, is low ($0.26$)
indicating that the $N_4$ term has disrupted the first Lyapunov vector and spread its energy to other Lyapunov vectors.
The fact that this disruption does not substantially alter
the time-mean energy extraction from the streamwise mean flow 
suggests that the time mean energetics resulting from projection on the Lyapunov vectors of 
$\U$  is not substantially altered by $N_4$ while the 
projection at an instant in time is (recall that in RNL $N_4$ is absent and this projection is onto the single top Lyapunov vector).  
To investigate this possibility we need to project the energetics on 
the Lyapunov vectors rather
than on the growth directions associated with the 
eigenvectors of $A+A^{\dagger}$, with $A$ the operator in 
\eqref{eq:RNSp1} governing the linear evolution of the perturbations about $\U$.
%
Note also that the decay rate of the  NL flow state due to dissipation  is  $0.0646 U_w/h$
 which is less then the corresponding decay rate of the first Lyapunov vector. This difference in  decay rate is
 explained by reference to  figure~\ref{fig:snap}, from which it is apparent that the  $n_x=1$ perturbation component of the
 NL  state is  of larger scale than the $n_x=1$ Lyapunov vector.
 Also, consider that, in the energetics
 of the $n_x=1$ perturbation component  in NL  there is a term not present in the corresponding Lyapunov vector:
 the energy interchanged with  the remaining $n_x\ne 0$ components, which is also shown in figure~\ref{fig:comp1}.
 The $n_x=1$ perturbation component of the NL state  exports energy to the other streamwise components of the flow
 and this nonlinear transfer contributes $0.0386 U_w/h$ at this wavenumber to the decay rate.
 This additional decay is just sufficient to reduce
 the  mean growth rate of the NL state
 to the expected value $\lambda_{state}=0$. 
 
 We conclude that the Lyapunov exponent of the fluctuating streamwise mean flow 
 $\U$  in NL turbulence  has been adjusted to near  neutrality which is consistent 
 with the parametric growth mechanism
 fully accounting for the maintenance of the 
 perturbation component of the turbulent state. The perturbation-perturbation nonlinearity, $N_4$,
 does not configure the perturbations to extract more energy from the mean flow  than they would in the absence of this term implying
 that  $N_4$
acts as
a disruption to the parametric growth process supporting the  
Lyapunov vector rather than augmenting the perturbation maintenance by the frequently hypothesized 
mechanism in which perturbation-perturbation nonlinearity  replenishes
 the subspace of perturbations configured to transfer energy from the mean flow to the perturbations.
 The fact that the mean NL flow has been adjusted to near neutrality indicates that the first Lyapunov vector should be a dominant component 
of the NL perturbation state. This will be examined in the next section.

\section{Analysis of perturbation energetics by projection onto  the Lyapunov vector basis}
\vskip0.2in

Despite the persuasive correspondence between the mean energetics of the NL 
perturbation state and the mean energetics of the top 
Lyapunov vector calculated using the associated fluctuating streamwise mean flow, $\U$, 
time series of perturbation growth rate for these shown in figure~\ref{fig:comp1}
and snapshots of the perturbation state and the Lyapunov vectors shown in figure~\ref{fig:snap}
reveal considerable differences. 
This suggests further analysis to clarify the relation between  the perturbation state and the Lyapunov vectors. 
The orthogonality property imposed on the Lyapunov vectors makes them an attractive basis for decomposing the
$n_x=1$ component of the NL perturbation state for the purpose of analyzing the relation between 
perturbation structure and energetics.  Expanding the Fourier amplitude of the $n_x=1$ NL perturbation state $\hat \u'$ in the basis of the orthonormal in energy $n_x=1$ Lyapunov vectors:  
\begin{equation}
\hat \u' (t) = \sum_i a_i(t) \u _{LV_i}(t),
\label{eq:expand}
\end{equation}
with  projection  coefficient:
\begin{equation}
a_i(t)=\left \langle \u _{LV_i}(t)\bcdot \hat \u' (t)\right \rangle_{x,y,z},
\end{equation}
we obtain that  the contribution to the energy of the perturbations accounted for by
projection of the perturbation state on Lyapunov vector $u_{LV_i}$ 
is $E_i=a^2_i(t)/2$.
The projection of the energy of the $n_x=1$ component of the perturbation state on the first 150 Lyapunov vectors is shown in figure~\ref{fig:proj}. 
The percentage of energy accounted for by projection on  the most unstable
Lyapunov vector
is $11\%$, significantly larger than the energy in each of the remaining Lyapunov  vectors. 
Adding the second unstable
Lyapunov vector raises this value to $17.4\%$ and the first 100
$n_x=1$ Lyapunov vectors account for $82 \%$ of the energy of the $n_x=1$ component { of the perturbation state as seen in figure~\ref{fig:cum}. 
In order to understand the significance of the Lyapunov vectors as a basis for representing the 
NL perturbation state we have determined the orthonormal structures  of the proper orthogonal decomposition  (PODs) of
the $n_x=1$ component of the NL flow  with the methods discussed in Ref.~\citep{Nikolaidis-etal-Madrid-2016}. Figure~\ref{fig:proj} and figure~\ref{fig:cum}
demonstrates that the Lyapunov vectors provide a good representation of the NL perturbation state. 
 We conclude that the energy of the perturbation state is partitioned into the Lyapunov vectors
 in the order of their Lyapunov exponent and that the Lyapunov vectors are a good basis to represent the perturbation turbulent state. 
While the energy accounted by the  POD basis is by construction 
expected to decrease monotonically (but not necessarily strictly monotonically) with the order  
of the POD this is not required of the Lyapunov vectors.  
Also note that the Lyapunov vectors are not constrained
by the optimality of the POD basis to be an inferior basis for spanning the energy, because the 
Lyapunov vectors are time dependent and could theoretically span all the perturbation
energy, as indeed is the case in RNL  for which the entire energetics is explained by the first Lyapunov vector.

In figure~\ref{fig:proj} we also show the average projection of the NL state on the
eigenvectors  of the time-mean operator $ A+A^{\dagger} $
 ordered in descending order of its eigenvalue. $A $ is the linear operator in 
\eqref{eq:RNSp1} governing the linear evolution of the perturbations about $\U$. The eigenvectors of $A+A^{\dagger}$
are the orthogonal directions associated with  instantaneous energy growth rate of perturbations  on 
the flow, $\U$.  The perturbation component has
 small projection on the first eigenvectors of $A+A^{\dagger}$ which are the 
 structures producing greatest  instantaneous energy growth rates. 
 The turbulent mean flow $\U$ is such that
 perturbations that lead to large instantaneous growth rate have large spanwise 
wavenumber and are located at the very high shear regions next to the
channel boundaries. Consequently, the state has a small projection 
on the structures producing the most rapid instantaneous growth and the turbulent perturbation component 
is concentrated on the directions with small positive instantaneous growth rate. Small projection of the perturbation state on the directions of maximum instantaneous growth rate was also seen in RNL simulations at $R=600$
\citep{Farrell-Ioannou-2017-sync}. What is remarkable and indicative of the fundamental 
role of the Lyapunov vectors in the dynamics of NL is the monotonic ordering 
of the perturbation energy in the Lyapunov basis; comparable ordering does not occur in 
the case of e.g. the dynamically important basis of the $A+A^\dagger$.


 In RNL simulations at $R=600$ the perturbation turbulent  state is
 entirely supported by the top Lyapunov vector and the energetics of the perturbation state
 consequently are the energetics of this single Lyapunov vector. The $N_4$ nonlinearity distributes  
 the perturbation energy over a subspace spanned primarily by the first few Lyapunov vectors, as shown in figure~\ref{fig:proj}.
 We can determine the distribution of the first $N$ Lyapunov vectors ordered in contribution to the perturbation state energy growth rate, $\dot E_{i, def}$ 
 by calculating
  \begin{equation}
\sum_{i=1}^N \dot E_{i , def} \equiv \left \langle \u_N' \bcdot \left ( -  \U  \bcdot \bnabla  \u _N' -
\u _N' \bcdot \bnabla  \U    \right ) \right  \rangle_{x,y,z,t}~,\label{eq:edefn}
\end{equation}  
where 
\begin{equation}
\u_N' (t) = \Re \left ( \sum_{\alpha=1}^N a_\alpha(t) \u _{LV_\alpha}(t) e^{ 2 \pi i x /L_x} \right ),
\end{equation}
 is the truncation of the representation of the $n_x=1$ perturbation state, given in~\eqref{eq:expand}, to the first $N$ Lyapunov vectors.
 From that calculation, we can then obtain the incremental contribution, $E_{i, def}$, of each Lyapunov vector. 
 We can similarly determine the contribution of each of the eigenvectors of $A+A^{\dagger}$ and of the PODs to the 
 energetics of the perturbation state.  
  The results, shown in Fig.  \ref{fig:edot}, reveal that the  Lyapunov vectors 
provide the primary support for
the perturbation energetics and their energetic contribution follows the Lyapunov vector growth rate ordering.
If the $N_4$ term were dominant in determining the structures supporting the perturbation state the energetics of the NL state 
would not be expected to so closely reflect the 
asymptotic structures of the Lyapunov vectors.    

 }

%

    
\section{Conclusions}
\vskip0.2in

Statistical state dynamics (SSD) based analysis of the transition to turbulence in Couette flow
 has revealed the sequence of 
 symmetry breaking bifurcations of the statistical state of the flow as the Reynolds number increases 
\citep{Farrell-Ioannou-2012, Farrell-Ioannou-2017-bifur}.
First the spanwise statistical symmetry of the laminar state is broken and later the temporal statistical symmetry is broken coincident with transition to the turbulent state.
Analyses made using SSD systems closed at second order have also  
 demonstrated that a realistic self-sustained turbulent state is maintained by  the parametric growth mechanism arising from 
 interaction  between the  temporally and spanwise spatially varying  streamwise-mean flow and 
 the associated perturbation component of the flow~\cite{Thomas-etal-2014,Thomas-etal-2015,Farrell-etal-2016-VLSM}.
  In these second order SSD systems the streamwise-mean flow 
 is necessarily adjusted   exactly to neutral stability, with the understanding that the time
 dependent streamwise mean flow is considered neutral when the first Lyapunov exponent is  zero.
 This result allows reinterpretation of the Malkus conjecture that the statistical state of inhomogeneous 
  turbulence has mean flow  adjusted  to neutral hydrodynamic stability.

  In this work the identification of the parametric mechanism supporting the perturbation component of turbulence  obtained using 
  statistical state dynamics in the RNL system has been  extended to NL.  
  While support of both the energy and energetics is on a single feedback neutralized Lyapunov vector 
  in the case of RNL, in  the case of NL  the energy and energetics are not confined to a single Lyapunov vector but 
  rather are spread by nonlinearity over the Lyapunov vectors. Importantly,  this support 
  of the perturbation structure and energetics  is ordered in the Lyapunov vectors 
  descending in   their associated exponents.
  The neutrality of the top Lyapunov vector in both RNL and NL, when account is taken of the transfer of energy to other 
  scales,   is interpreted as implying that the Malkus conjecture is valid if neutrality of the mean state is 
  interpreted as neutrality of the top Lyapunov vector(s). Consistent with the parametric 
  mechanism sustaining the turbulence, 
   the perturbation structure is concentrated on
  the top Lyapunov vectors of the time varying streamwise-mean flow and ordered in their Lyapunov exponents.
  Identification  of the dynamical support of RNL and NL turbulence by the marginally stable Lyapunov 
  vectors with associated parametric growth mechanism vindicates the conjecture 
  that the dynamically relevant mean flow in turbulent shear flow  
   is not the  streamwise-spanwise-temporal mean flow but rather the time and spanwise varying  streamwise-mean flow 
   which incorporates   the statistical  symmetries obtained from study of second order closures of the statistical state dynamics of wall-bounded turbulence.
 The perturbation-perturbation nonlinearity, 
  does not configure the perturbations to extract more energy from the mean flow  
 than they would in the absence of this term implying
 that  the nonlinearity  acts as
a disruption to the parametric growth process supporting the  perturbation field
 rather than augmenting the perturbation maintenance by the frequently hypothesized 
mechanism in which perturbation-perturbation nonlinearity  replenishes
 the subspace of perturbations configured to transfer energy from the mean flow.
  The fact that the mean NL flow has been adjusted to Lyapunov
  neutrality and that the Lyapunov vectors support both the energy and the energetics of the perturbation component of the turbulent state   
  indicates that the parametric growth mechanism on the fluctuating 
  streamwise mean flow and its regulation by Reynolds stress 
  feedback which has been identified to support RNL turbulence is also the mechanism underlying the support of NL turbulence.

  
\ack
This work was funded in part by the Coturb program of the European Research Council. We thank Javier Jimenez for his reviewing comments. Marios-Andreas Nikolaidis acknowledges the support of the Hellenic Foundation for Research and Innovation (HFRI) and the General Secretariat for Research and Technology (GSRT).  Brian F. Farrell was partially supported by the U.S. National Science Foundation under Grant Nos. NSF~AGS-1246929 and NSF~AGS-1640989.

\providecommand{\newblock}{}

\end{document}